\RequirePackage{fix-cm}
\documentclass{svjour3}                     
\smartqed  
\usepackage{graphicx}
\usepackage{amsmath,amssymb,amsfonts}
\usepackage{algorithmic}
\usepackage{graphicx}
\usepackage{textcomp}
\usepackage[usenames,dvipsnames,table]{xcolor}

\definecolor{gray20}{gray}{0.8}
\definecolor{gray5}{gray}{0.95}
\definecolor{olivegreen30}{RGB}{155,187,89}	

\usepackage{float}
\usepackage{subfigure}
\usepackage{soul}    
\usepackage{multirow}

\usepackage{pgfplots}
\usepackage{tikz}
\usetikzlibrary{pgfplots.dateplot}
\pgfplotsset{compat=1.17}

\usepackage{url}
\usepackage{orcidlink}

\usepackage{hyperref}

\begin{document}

\title{Incidental Data: Observation of Privacy Compromising Data on Social Media Platforms
}

\titlerunning{Incidental Data: Observation of Privacy Compromising Data on Social Media}        


\author{Stefan Kutschera \orcidlink{0000-0002-7547-5137}}

\institute{Stefan Kutschera \at
              Graz University of Technology\\
			  Institute of Software Technology, Inffeldgasse 16b, 8010 Graz, AUSTRIA \\
              \email{stefan.kutschera@ist.tugraz.at} }

\date{Received: date / Accepted: date}

\maketitle


\begin{abstract} 
Social media plays an important role for a vast majority in one's internet life. Likewise, sharing, publishing and posting content through social media became nearly effortless. This unleashes new threats as unintentionally shared information may be used against oneself or beloved ones.
With open source intelligence data and methods, we show how unindented published data can be revealed and further analyze possibilities that can potentially compromise one's privacy. This is backed up by a popular view from interviewed experts from various fields of expertise.
We were able to show that only 2 hours of manually fetching data are sufficient in order to unveil private personal information that was not intended to be published by the person. Two distinctive methods are described with several approaches. From our results, we were able to describe a thirteen-step awareness guideline and proposed a change of law within Austrian legislation.
Our work has shown that awareness among persons on social media needs
to be raised. Critically reflecting on our work has revealed several ethical implications that made countermeasures necessary; however, it can be assumed that criminals do not do that.

\end{abstract}

\section{Introduction}\label{chap:intro}

\subsection{Problem statement}\label{sec:problem}
As more and more people become connected to the internet, the obstacles of publishing own created content become more uncomplicated. Likewise, more individuals can consume the published content. 
The simplicity of a connected world has heavy implications for privacy, security, and safety. Not everyone who contributes content has the time or awareness to cross-check for any unwanted data included in the published social media posting.
Besides the desired information a posting should transfer, the unaware and unwanted published data may be used to gather even more information, for instance, previous owners, cost of recent renovations, property tax, telephone number, and date of birth, to name a few.

Unwanted data within a public social media posting may be used for targeted attacks as criminals can hide behind their computers and gather information that would have needed careful physical observation in front of a victim's location. Such possible attacks might include but are not limited to burglary, phishing, or identity theft.

\subsection{Incidental data}\label{sec:incidentaldata}
The term \emph{incidental} is used in the context of a medical examination, it means something fortuitous has been found; thus \emph{``incidental findings''}. \cite{Wolf2:2008} However, \emph{incidental} was also used to describe unanticipated observations during an on-premise information security audit. \cite{Casanovas:2019} For example, an x-ray of a broken chest reveals a suspicious tumorous tissue within the lungs, or looking behind a server rack reveals a wet spot indicating a water leak. 
Taking these points into consideration\emph{``incidental data''} in our work and beyond shall be known as unintentional hidden information in a public posting on the internet or on social media platforms. \cite{kutschera:2021} The following situation could be an example: a person wishes to show how red, healthy and tasty the tomatoes are growing in their own garden with a video or a photo but unknowingly reveals the residential address of the own home and also a great deal more private information.

\subsection{Research questions}\label{sec:researchquestions}
The following research questions will be answered in this context as part of the underlying research conducted as part of the author's master thesis. \cite{kutschera:2021}
\begin{enumerate}
    \item What is the legal aspect of unveiling personal information found in videos? 
    \item To what extent is the information that has been gathered of value to a person or an organization with criminal intentions?
    \item What methods can extract incidental data in a maximum of 2 hours using OSINT data?
\end{enumerate}

\subsection{Hypothesis}\label{sec:hypothesis}
Postings that are made freely available on social media by a person, frequently include incidental data which can be accessed and also used to harm their author. 

\paragraph{Null-Hypothesis}
Out of an arbitrary small selection of public social media accounts, none include compromising private information \emph{incidental data}.

\subsection{Methodology}\label{sec:method} 
Open-source intelligence (OSINT) data and methods are used to gather and fetch data from various social media platforms manually. After allocating personal data that may be posted unintentionally, this data is used to extract further information from freely available information or information services. In order to have comparable results and not extensively fetch data from a single profile, the gathering process should not exceed 2 hours. This threshold is also a countermeasure to prevent allegations of stalking. The popular view from experts will be obtained from interviews and analyzed using qualitative content analysis. \cite{MayringPhilipp2010QI:G}
The incidental data can subsequently be evaluated to assess its economic value from a criminal perspective. In addition, a fictitious scenario based on this work will reveal the legal aspect of fetching incidental data in the course of a case law survey.

\subsubsection{Ethic \& morality }\label{subsec:ethics}
Personal data is a precious asset in the world of today, not only for individuals but also for companies and in both cases, criminals can be at work using personal data in an attempt to harm people. \cite{ablon:2018}
It became evident in the course the research that our work brings a considerable degree of responsibility with it and that an evaluation of the ethical aspect needs to be carefully considered. In this context, we decided to undergo an ethics self-assessment test. \cite{erc:2018}
The ethics self-assessment test consists of 20 main questions, where an answered ``Yes'' implicitly or directly indicates a particular risk. As Table~\ref{yesnoratio} shows, our work needed to be answered in 9, thus 45\% of all 20 main questions with yes. Coupled with the questionnaire, measures that counteract certain specific risks are also mentioned. We introduced specific measures derived from the feedback of the ethics self-assessment test. For example, limit the number of hours spent on fetching incidental data. Furthermore, in the case of persons without a professional security and privacy background, ask for permission to fetch incidental data from the social media profile in question.  

\begin{table}[h]
    \caption{This table gives an overview of how we answered the questions from the ethical self-assessment test and how the overall ratio of Yes/No is. As in this case, the ratio is 9/11.}
    \begin{tabular}{l|llllllllll|l}
        Question \# & 1 & 2 & 3 & 4 & 5 & 6 & 7 & 8 & 9 & 10 &           \\
        \hline
        Questions   & 3 & 2 & 1 & 2 & 1 & 5 & 3 & 1 & 1 & 1  & $\sum 20$ \\
        YES         & 0 & 1 & 0 & 2 & 0 & 2 & 2 & 1 & 1 & 0  & $\sum 9$  \\
        NO          & 3 & 1 & 1 & 0 & 1 & 3 & 1 & 0 & 0 & 1  & $\sum 11$ \\ 
        \hline
    \end{tabular}
    \label{yesnoratio}
\end{table}

\section{Related work}\label{chap:related}

\subsection{Open source intelligence - OSINT}\label{sec:osint}
The process by which intelligence is produced and collected from publicly available information and distributed to persons with specific intelligence requirements is specified as OSINT. For example, an attorney or a civil engineer using OSINT would mean official documents would be in use, whereas, for a civil foreign intelligence service, OSINT would be a collection of broadcast news from a foreign country. Moreover, creating reports with the gathered information is also covered within the term OSINT due to the risk of information becoming unavailable. \cite{bazzel2021} 

When OSINT information is collected, a reproducible procedure cannot be drawn in general. Despite this fact, a generalization of specific steps can be made as discussed in Table~\ref{osinttable}

\begin{table*}
\caption{Tabular overview of abstract path during OSINT collection intelligence. \cite[p.606]{bazzel2021} }
    \begin{tabular}{lp{0.2\columnwidth}|p{0.65\columnwidth}}
    \hline
    \rowcolor{gray5}\# & Step & Description \\
    \hline
    1  & Triage                               & Creation of mission overview. Definition of expectation, questions.\\
    2  & Preparation of Tools                 & Programs, tools, and services that are needed to support and answer the Triage step are made ready for usage. \\
    3  & Closed Source Data Queries           & (Optional) Databases locked through either payment or governmental access are queried against. \\
    4  & OSINT: Query All Known   Identifiers & Necessary and available databases, -sources and techniques are executed.\\ 
    5  & Collection                           & The leads serve as an input for the collection or as new evidence that can be used again in step 4. \\
    6  & Analysis                             & All gathered data is used to depict ``bigger picture'' and provide a good overview \\
    7  & Reporting                            & All gathered data is used to answer definitions from \emph{Triage} within a final report. \\
    8  & Cleanup/Archiving                    & Virtual machines are either reverted and archived to provide an unbiased starting point. \\
    \end{tabular}
    \label{osinttable}
\end{table*}

\subsection{Value of personal data}\label{sec:data_worth}
OSINT information adds by its definition a certain value to the conducted investigation. In addition to this, personal, financial or health data may be used to gain an advantage, to profit or to harm others by an intruder. \cite{ablon:2018,bazzel2021}
Moreover, a simple identity mistake was made in the case of \textit{David Quintavalle}, and he was later proved innocent by the Federal Bureau of Investigation (FBI). \cite[15:19min]{bazzelpodcast214:2021}

As made clear by Heimo Flechl, BA MA, Head of the Unit of OSINT \& Crime Trends within the Criminal Intelligence Service Austria, gathered incidental data is worth very little when offered for sale on the dark web. Nevertheless, this situation can change when a method such as "Crime as a Service" is used or requested. "Crime as a Service" may, in this particular case, describe the information gathering process on a contractual basis. In such a case, the value of this illicitly gathered information can range in value from under a hundred EUR to several thousand EUR. Another aspect could be not to sell but to exploit the information to harm a person in numerous ways or   occasions such as causing repair costs after a forceful entry, bringing about loss of reputation, or the loss of a job. Whatever the case, however, cash is still in favour among criminals. \cite{Weber2019,kutschera:2021,hessen:2020,hagenplz:2021}

\subsubsection{Public versus social watchdogs}\label{subsec:publicwatchdogs}
A phenomenon of the digital revolution was described in an analysis of case law on Art.~10~ECHR (freedom of expression). It was found during an analysis of case law on Art.~10~ECHR (freedom of expression) that certain professional groups enjoy even greater protection or freedom are divided into two groups, so-called "public-watchdog" and newly found so-called "social watchdogs". Whereas public-watchdogs are journalism in the classic sense, made available to the general public in print or digital broadcast media, social-watchdogs are citizens without a professional journalistic background. They must have a similar broadcast range compared to their shared content and opinion on social media with public-watchdogs. However, current legislation grants public-watchdogs more freedom. Furthermore, the standard of due diligence is no higher for social-watchdogs than it is to their professional counterparts, the public-watchdogs. \cite{holoubek:2016}

\subsection{Human Rights Convention}\label{sec:humanrights} %
Since human rights are the most fundamental rights any  person can have, their inviolability would appear to be an obvious requirement. However, those rights are not absolute but tend to interfere with each other in such a way that clear borders cannot be drawn. This interference is especially true for \emph{Art.8~ECHR the right to respect for private and family life} and \emph{Art.10~ECHR the freedom of expression} as shown as follows.

With a look at case law, it becomes clear that even the private life of publicly well-known people is protected. The case \textit{Alkaya~v.~Turkey~(42811/06~ECHR)}, and its ruling supports this, as a well-known film and theatre actress from Turkey had a home robbery while at home, and a national newspaper later published pictures and information on her whereabouts. Further, the ruling of \textit{Satakunnan~Markkinapörssi~Oy~and~Satamedia~v.~Finland (931/13 ECHR)} the ECHR stated in the court assessment ``134.\ The fact that information is already in the public domain will not necessarily remove the protection of Article 8 of the Convention.''
In contrast, the ruling of \textit{Cengiz~and~others~v.~Turkey} shows that social media platforms are indispensable tools for exercising the right of freedom of expression. The case of \textit{Von~Hannover~v.~Germany} clearly hints that Art.~10~ECHR and Art.~8~ECHR have to be treated with equal respect. In the matter that several photographs were taken and published in a newspaper the case did not withstand the balancing between \emph{Art.10~ECHR, the freedom of expression}, and \emph{Art.8~ECHR, the right to respect for private and family life}.

\section{Implementation} \label{ch:implementation}

\subsection{Manual Gathering Example \#1}\label{subsec:gatehringprocessyoutube}
\begin{figure}[h]
    \centering
    \begin{minipage}{\columnwidth}
         \subfigure[Left side of surroundings.\cite{liveeachday_1:2018} ]{\label{PIC_LiveEachDay_backyard_surrounding_1}\includegraphics[width=0.32\columnwidth]{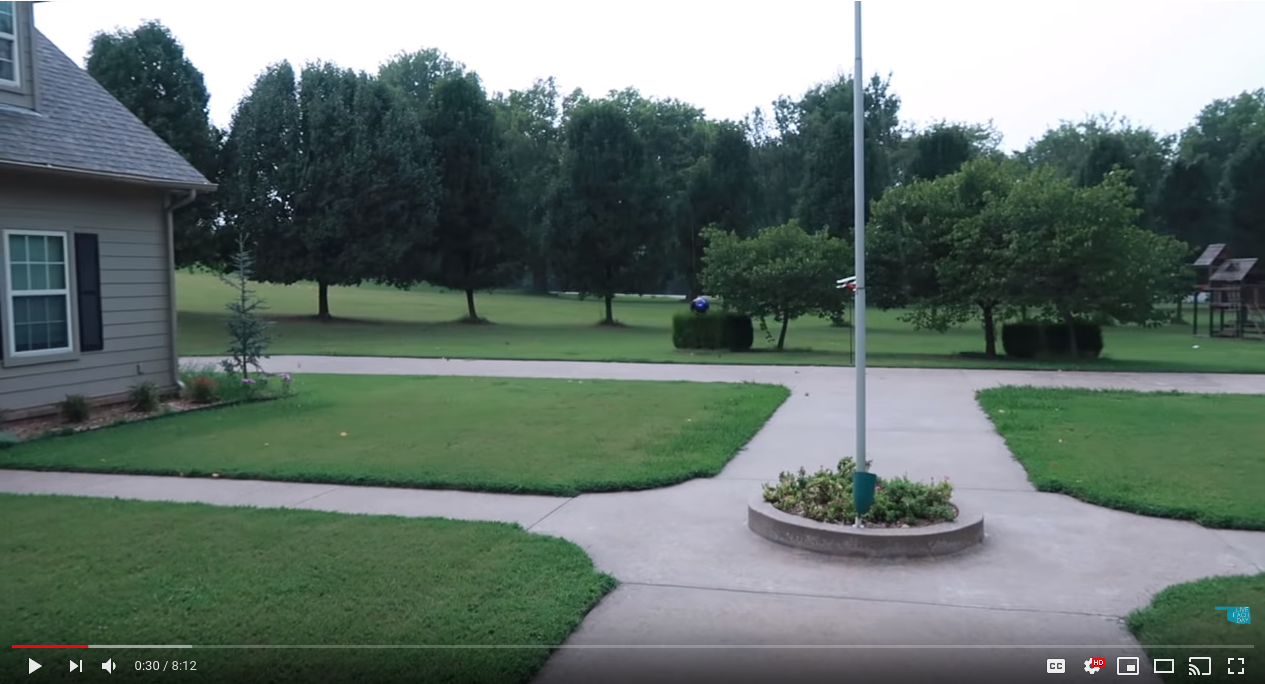}}
        \subfigure[Right side of surroundings. \cite{liveeachday_1:2018} ]{\label{PIC_LiveEachDay_backyard_surrounding_2}\includegraphics[width=0.32\columnwidth]{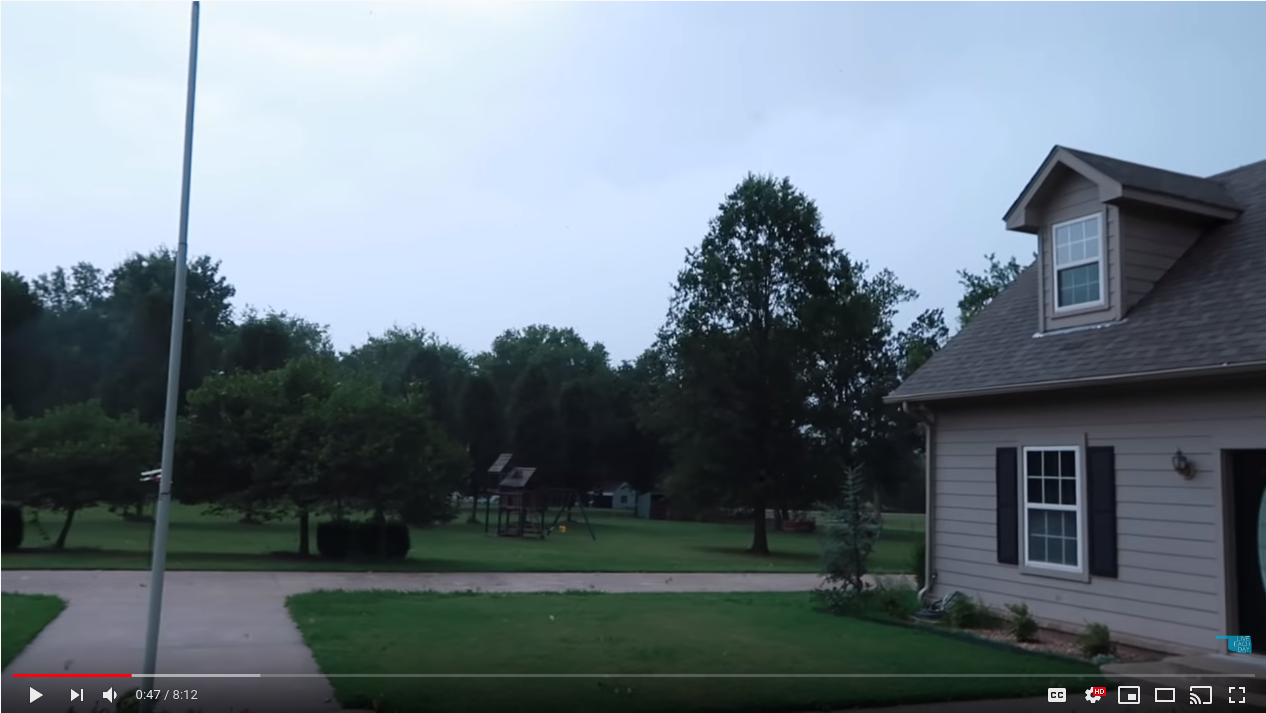}}
        \subfigure[Smartphone revealing the GPS location. \cite{liveeachday_1:2018} ]{\label{PIC_LiveEachDay_gps_location}\includegraphics[width=0.32\columnwidth]{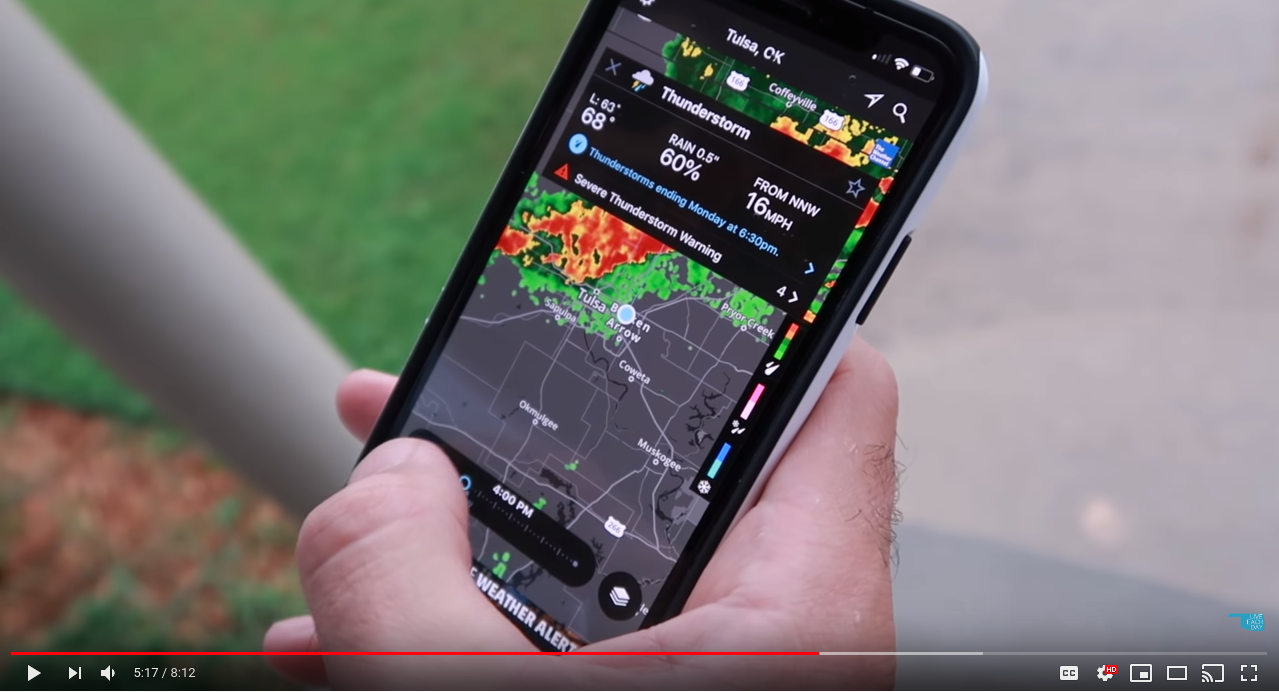}}
    \end{minipage}
    \caption{Depicts the unmodified screenshots taken from the YouTube video called \emph{``Wild Oklahoma Weather''} who are later used to identify the physical location.}
\end{figure}

In the video \emph{``Wild Oklahoma Weather''} from the YouTube channel\emph{``LiveEachDay''} a backyard scene reveals information on the surroundings of a private home as, for instance, a specific U-shape of the building with a US flag in the middle of the backyard. Albeit, it would be a lengthy process to search for the whole state of Oklahoma, as the title indicates. Notably, the scene, where a map from the rain radar within a weather app is shown to the camera as depicted in Figure~\ref{PIC_LiveEachDay_gps_location}, reveals the GPS position. 

\begin{figure}[h]
    \centering
    \begin{minipage}{\columnwidth}
        \subfigure[Zoomed and rotated snipped of the map and GPS marker visible on the smartphone from screenshot visible in Figure~\ref{PIC_LiveEachDay_gps_location}. \cite{liveeachday_1:2018}]{\label{PIC_liveeachday_mobile_map}\includegraphics[width=0.32\columnwidth]{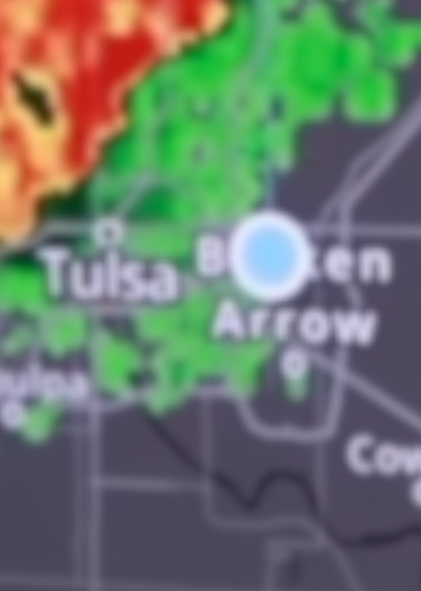}}
        \subfigure[Showing the same city fragment of Tulsa, Oklahoma, USA on Google Maps as visible in the screenshot from Figure~\ref{PIC_liveeachday_mobile_map}. \cite{googletulsa:2019}]{\label{PIC_liveeachday_google_map}\includegraphics[width=0.32\columnwidth]{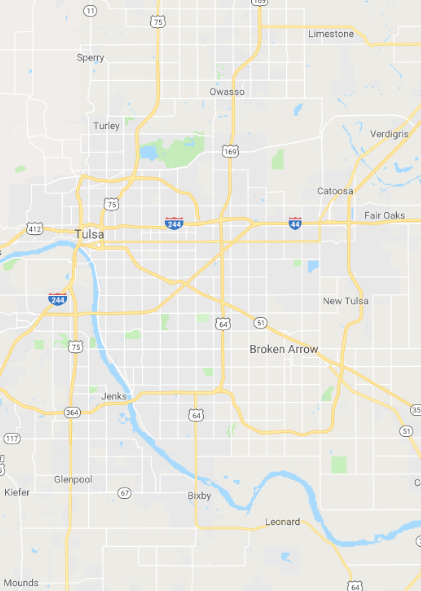}}
        \subfigure[Utilizing the distinct street arrangements a morphed images was created revealing a usable search area on the map. \cite{googletulsa:2019,liveeachday_1:2018}]{\label{PIC_liveeachday_map_overlay}\includegraphics[width=0.32\columnwidth]{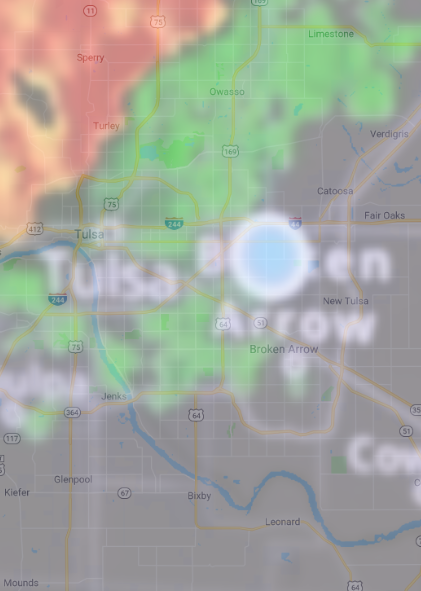}}
    \end{minipage}
    \caption{Depicts the morphing process of a screenshot with a map provider as a second data source.}
\end{figure}

With this information, the search radius can be drastically reduced as shown in Figure~\ref{PIC_liveeachday_mobile_map} - \ref{PIC_liveeachday_map_overlay} using a zoomed, rotated and cropped image as shown in Figure~\ref{PIC_liveeachday_mobile_map}). Figure~\ref{PIC_liveeachday_mobile_map}) is eventually used to morph and match the corresponding map from Google Maps as shown in Figure~\ref{PIC_liveeachday_map_overlay}).

\begin{figure}
    \centering
    \includegraphics[width=\columnwidth]{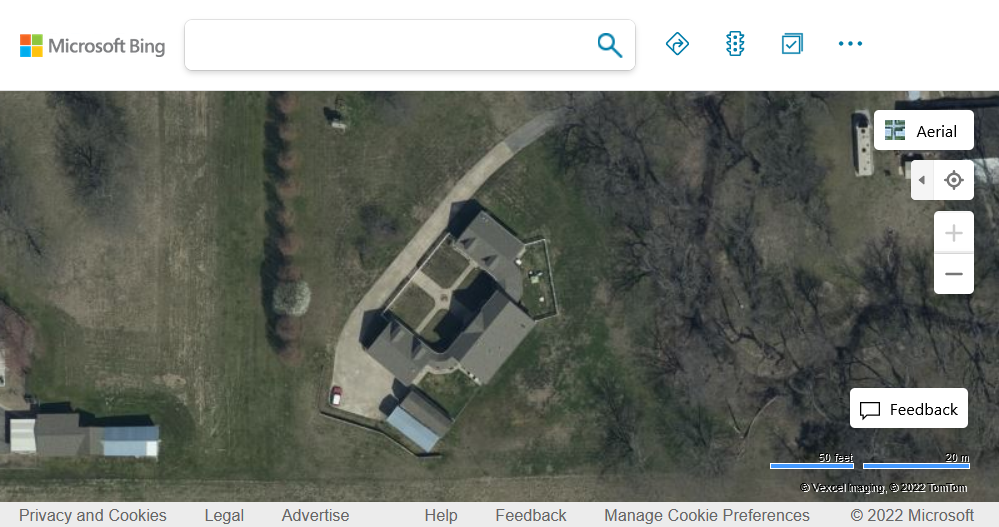}
    \caption{Screenshot from Bing Maps of the search result for a physical location after analysing satellite images for features as shown in Figure~\ref{PIC_LiveEachDay_backyard_surrounding_1}-\ref{PIC_LiveEachDay_backyard_surrounding_2}. \cite{bingmap:2019}}
    \label{PIC_LiveEachDay_second_finding}
\end{figure}

Since the search radius for the home was drastically reduced due to the GPS location provided and eventually the address was discovered within 10 minutes using satellite images, as seen in Figure~\ref{PIC_LiveEachDay_second_finding}. By using the address as an additional search input, it was possible to find information on the foreclosed previous owners. \cite{foreclosure:2019} In this foreclosure details, ultimately, the real estate parcel number ``9941****290'' allowed us to query the public register of Tulsa City unveil information about the previous and current owners, with the implication of these being members of a close family in the listed current and previous owners with the inclusion of tax information, home improvements and a list of further documents and pictures as seen in Figure~\ref{PLOT_table_size}-\ref{fig:tulsacountyregister}. The evidence proved consistent as the information from the public register of Tulsa City matched the foreclosure information. 

\pgfplotstableread{listing_price.dat}\stats
\begin{figure}[H]
    \centering
        \begin{minipage}{0.45\columnwidth}
            \begin{tikzpicture}[scale=0.55, transform shape]
                \begin{axis}[date coordinates in=x,ylabel=Property Price in USD, xticklabel style={rotate=45},
                scaled y ticks=false,
                y tick style={/pgf/number format/fixed},
                y label style={at={(axis description cs:-0.225,.5)},anchor=south},
                legend pos=outer north east]
                \addplot [smooth,mark=*,blue] table	[x={Date},y={Price}]{\stats};
                \end{axis}
            \end{tikzpicture}\caption{Diagram of prices from June 2016 to June 2017, where the the owner of the YouTube channel LiveEachDay allegedly bought the property. \cite{trulia:2019,zillow:2019}}
            \label{PLOT_table_size}
        \end{minipage}
        \begin{minipage}{0.45\columnwidth}
            \includegraphics[width=1\columnwidth]{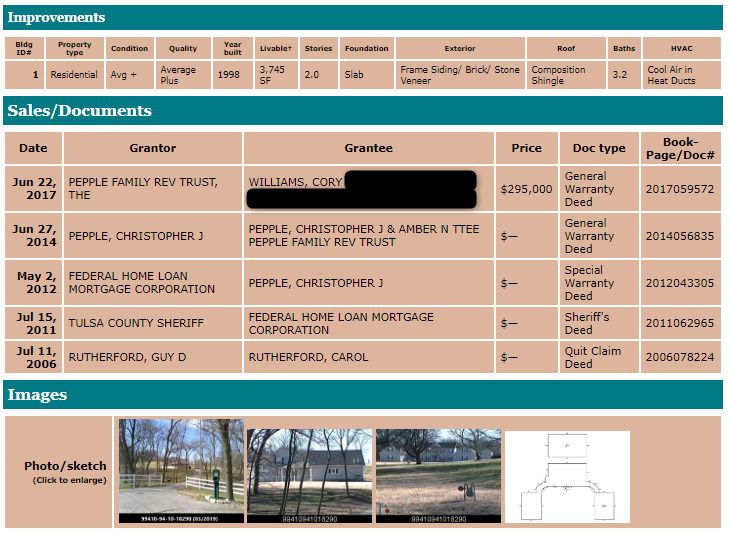}
            \caption{Excerpt from the public register of Tulsa City showing sales on June 2017 to the owner of the YouTube channel LiveEachDay. \cite{tulsacounty:2019}}
            \label{fig:tulsacountyregister}
        \end{minipage}
\end{figure}

\subsection{Manual Gathering Example \#2}\label{sec:troyhuntmanualgather}
In the following example, posts on the Twitter profile of Mr.~Troy Hunt, a public figure known as an Australian web security consultant, outreach on security topics, and creator of \emph{``Have I Been Pwned''}, are observed for indications on incidental data. 

\begin{figure}[H]
    \centering
    \subfigure[Showing car in front of alleged home. \cite{hunt_tw_piccar:2020}]{\label{fig:twitter_car}\includegraphics[width=0.32\columnwidth]{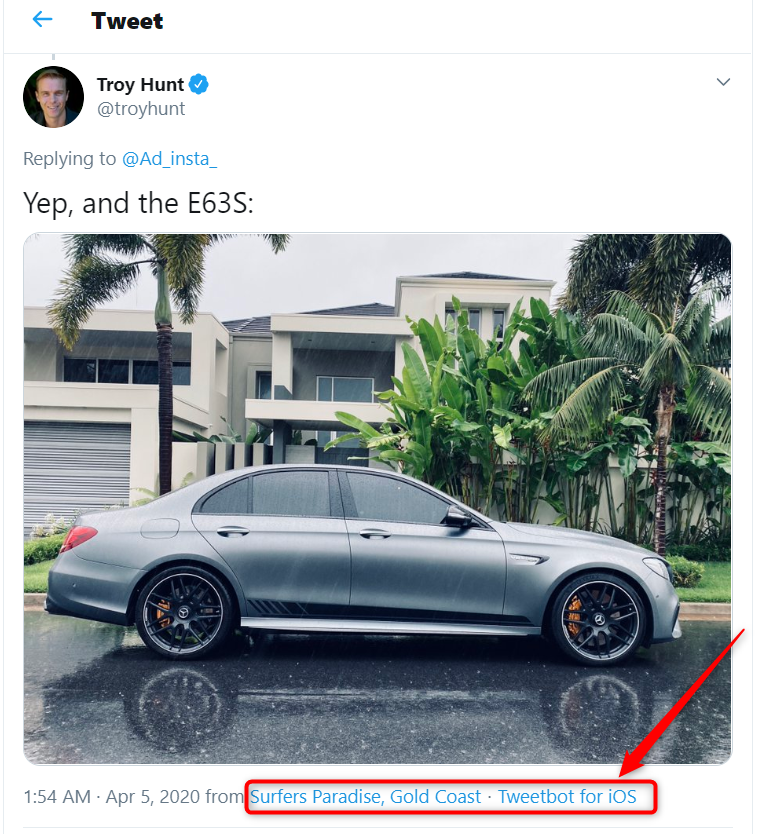}}
     \subfigure[Alleged backyard showing the sea view. \cite{hunt_tw_picsea:2020}]{\label{fig:twitter_th_sea}\includegraphics[width=0.32\columnwidth]{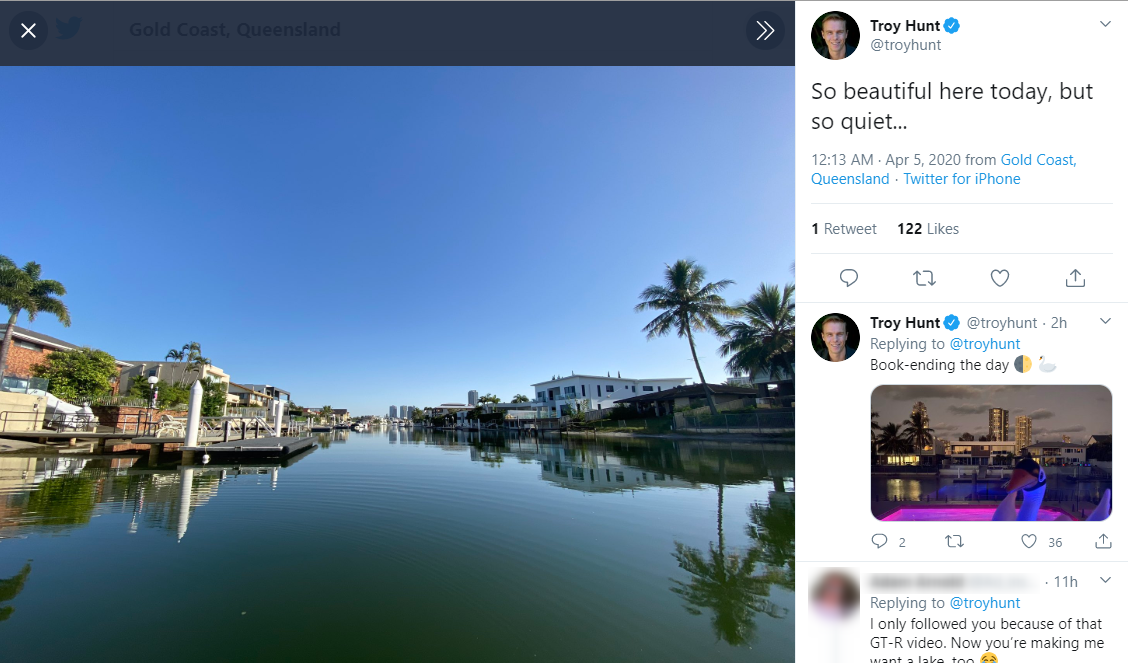}}
      \subfigure[Alleged backyard showing a pool. \cite{hunt_tw_picpool:2020}]{\label{fig:twitter_th_whirlpool}\includegraphics[width=0.32\columnwidth]{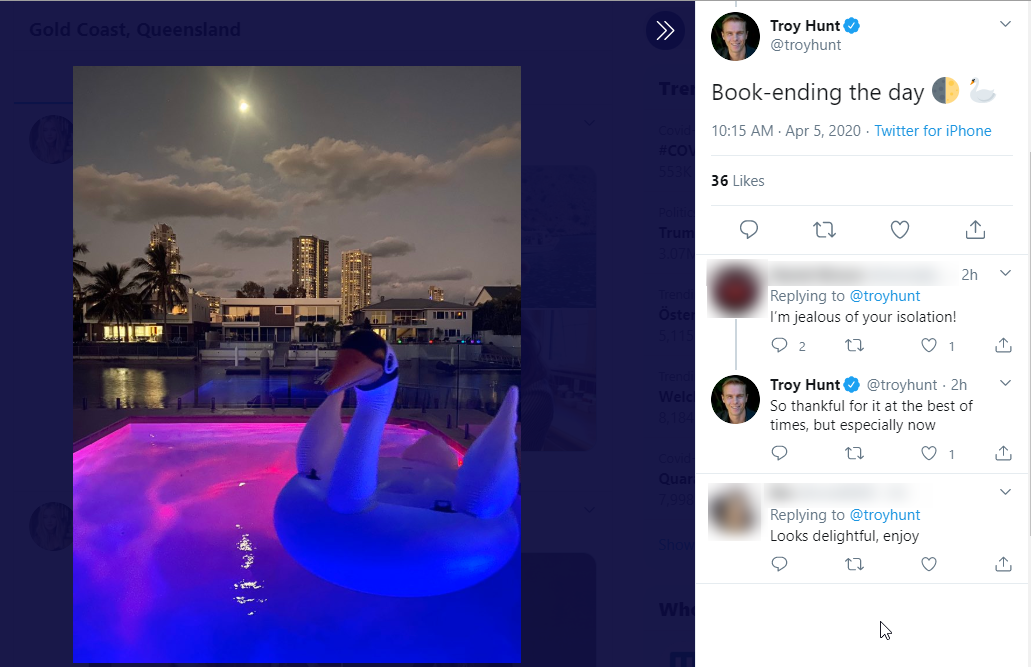}}
    \caption{Showing several posts of Mr.~Hunt depicting images from the alleged front respectively backyard of his home.}
\end{figure}

In a tweet, Mr.~Hunt was asked if he still owns his Mercedes car; he replied with an image showing the car in front of a house. With the implied assumption that the house that can be seen here might well be his home, an OSINT search, including posted images from his backyard, was used, see Figure~\ref{fig:twitter_car}-\ref{fig:twitter_th_whirlpool}. Looking for the striking skyscraper pattern in the background using the Google Maps 3D-view, an address was found within 25min.
Moreover, by utilizing the address, it was possible to gather information on alleged relatives, property price, size, date of purchase and telephone numbers of Mr.~Hunt and his alleged relative as visible but redacted in Table~\ref{tab:exampledatatroyhunt}. Further, the telephone number was also linked to messenger apps such as Signal and Telegram, where the profile picture depicting Mr.~Hunt was visible.

\begin{table}[h]
    \caption{Shows the redacted information collected from various public free available sources.}
    \begin{tabular}{p{0.30\columnwidth}|p{0.60\columnwidth}}
        \hline
        \rowcolor{gray5}\multicolumn{2}{c}{Overview} \\
        \hline
        Public channel     & Troy Hunt     \\
        Full name     & Troy Hunt     \\
        Birthday     & ---     \\
        Address     & 5**** A**** D****, 4217, Surfers Paradise, Queensland, AUSTRALIA     \\
        Relative     & K**** Hunt   \\
        Additional Information     & Property Lot **** on RP****, 721 sqm     \\
        Additional Information     & Property bought on 23rd March 2018 for **** AUD    \\
        Additional Information     & Phone (Troy Hunt): +61****76      \\
        Additional Information     & Phone (K**** Hunt): +61****88      \\
        \hline
    \end{tabular}
    \label{tab:exampledatatroyhunt}
\end{table}

\begin{figure}
    \centering
    \begin{minipage}{0.45\columnwidth}
        \subfigure[Shows a GPS tracked route within the Yulara, Northern Territory area and implicitly confirms a physical presence. \cite{th_tw_dateconfirmation:2021}]{\label{PIC_th_tw_confirmation}\includegraphics[width=0.99\columnwidth]{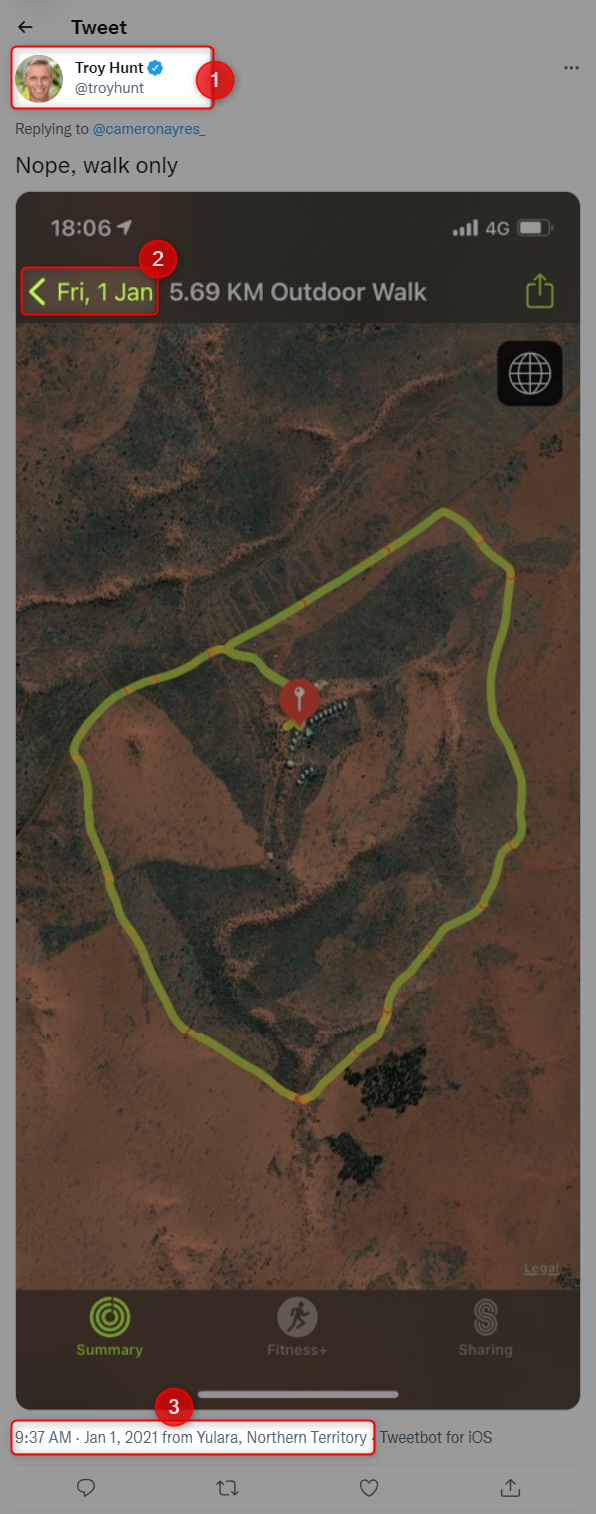}}
    \end{minipage}
    \begin{minipage}{0.45\columnwidth}
        \subfigure[Announcement of an upcoming road-trip through Australia including the route as a Google Maps link. \cite{hunt_tw_pic3:2020}]{\label{PIC_th_roadtrip_1}\includegraphics[width=0.99\columnwidth]{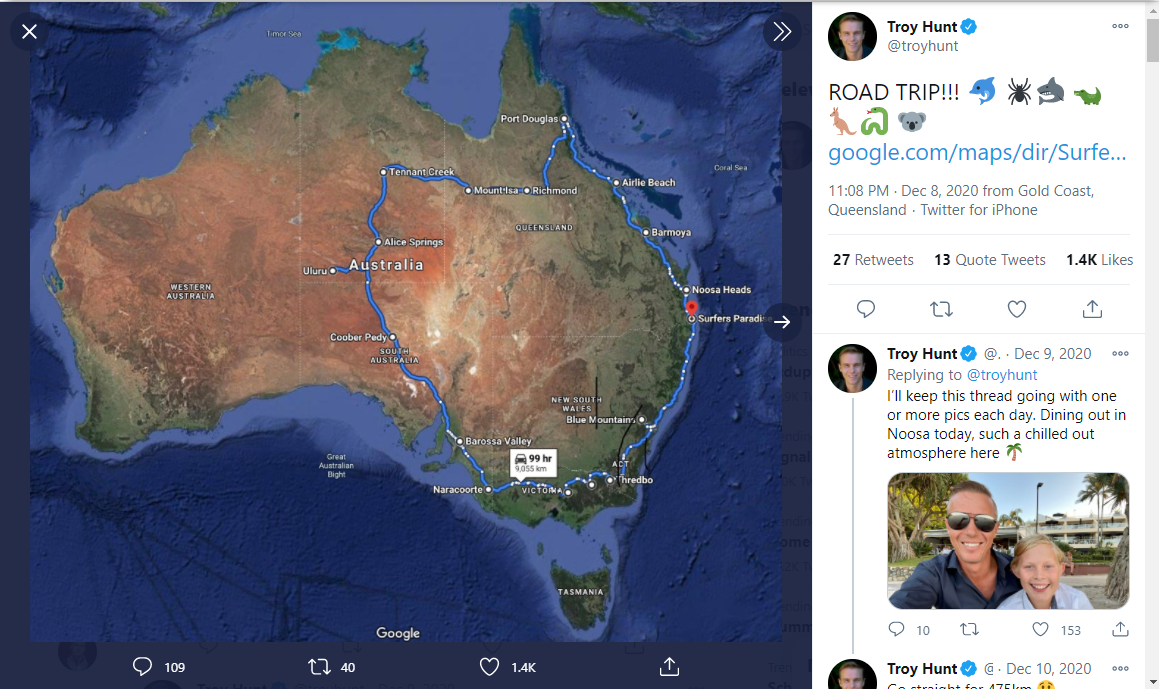}}
        \subfigure[Depicts on-time updates from a location that matches the itinerary. \cite{hunt_tw_pic2:2020}]{\label{hunt_tw_pic2}\includegraphics[width=0.99\columnwidth]{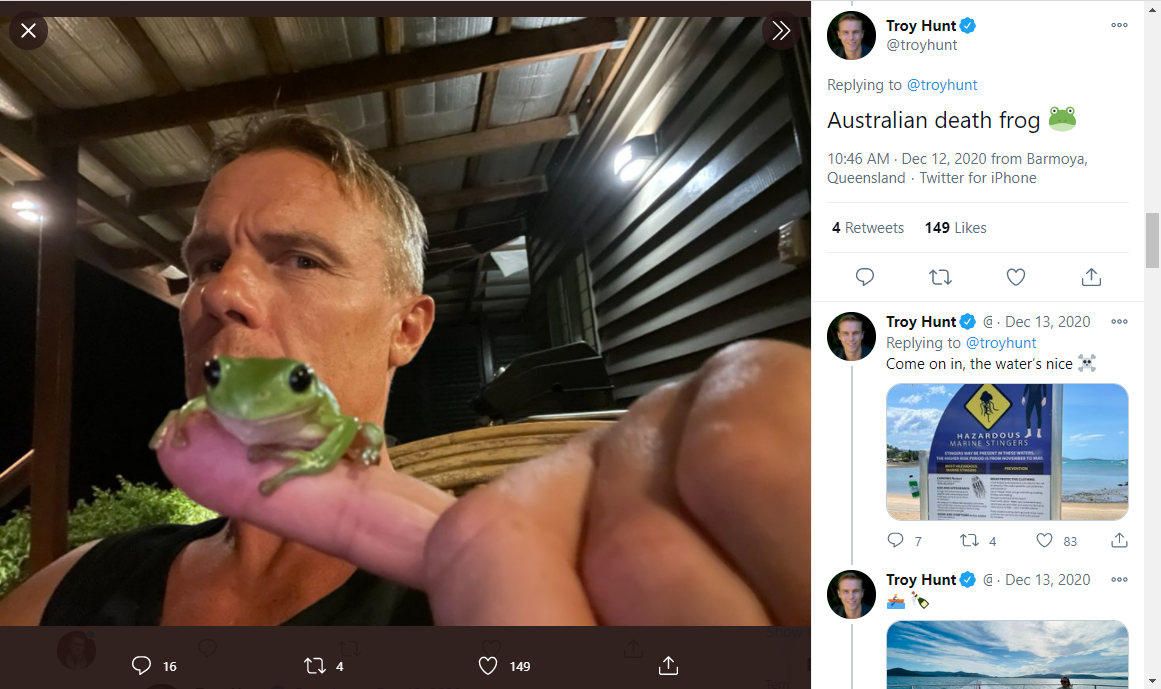}}
        \subfigure[Depicts a livestream on YouTube where streaming-time and -location match and strengthen the evidence. \cite{th_yt_date_confirmation:2021}]{\label{th_yt_date_confirmation}\includegraphics[width=0.99\columnwidth]{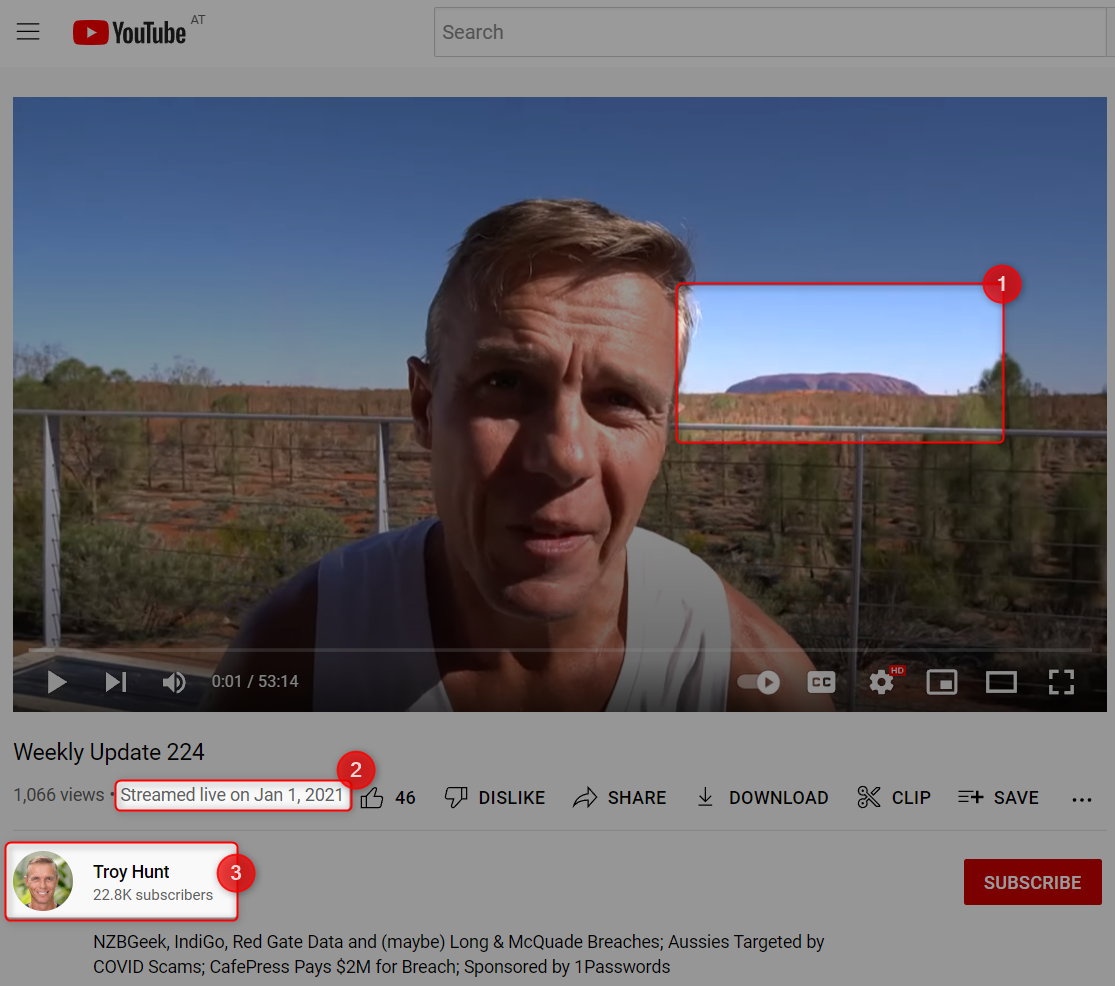}}
   \end{minipage}
   \caption{Screenshots depicting the vacation announcement and on-time updates of current position.}
\end{figure}

The most surprising fact was the announcement of an upcoming adventure respectively vacation with a detailed itinerary of around 9\,000 km and later updates from various locations during his ongoing trip, as seen in Figure~\ref{PIC_th_tw_confirmation}-\ref{th_yt_date_confirmation}. Within Figure~\ref{PIC_th_roadtrip_1} a detailed itinerary shows feature destination ahead of time. Evidence implicitly confirms his physical presence later in the Yulara area within the Northern Territory. At first glance, Figure~\ref{PIC_th_tw_confirmation} has inconsistency with respect to screenshot time and publishing time, as the screenshot time is clearly ahead of the posting time. However, it turned out that posting times on Twitter match the time zone of the viewer. When the time zone is manually changed to UTC+9:30, matching Northern Territory time zone, the posting time was 18:07, thus 1 minute after the posted screenshot on Twitter was taken, see Figure~\ref{PIC_th_tw_confirmation}. The evidence found strongly hints that the adventure trip was indeed posted ahead of time and with current updates of different locations.

\begin{figure}[h]
    \centering
    \includegraphics[width=0.5\columnwidth]{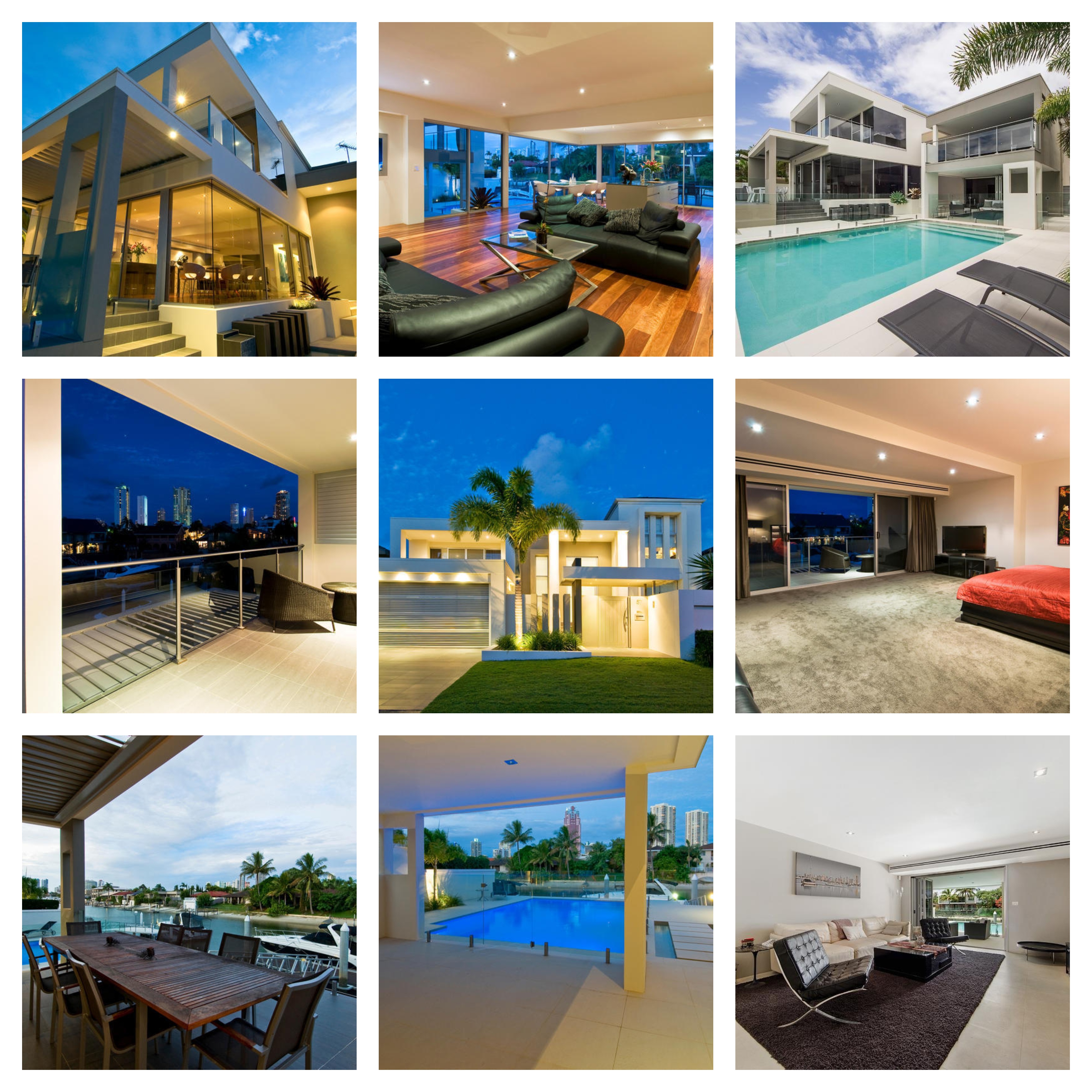}
    \caption{A collage with 9 out of 30 pictures from the alleged home of Mr.~Troy Hunt. \cite{oth:2020}}
    \label{fig:pic_collage_troyhunt_home}
\end{figure}

\subsection{Off-the-Grid}\label{sec:offthegrid}
We made a relatively quick setup to show how quick and easy it is to hide one's identity and go, so-to-speak \emph{off-the-grid}. Assuming that the internet service provider (ISP) has the technical ability to track our online behavior and legal obligation to hand over information to authorities and further, the fact that the the-onion-routing (TOR) browser has publicly known entry- and exit nodes, the ISP shall not be able to track that we are entering into a TOR node.\par
\begin{figure}[H]
    \centering
    \includegraphics[width=\columnwidth]{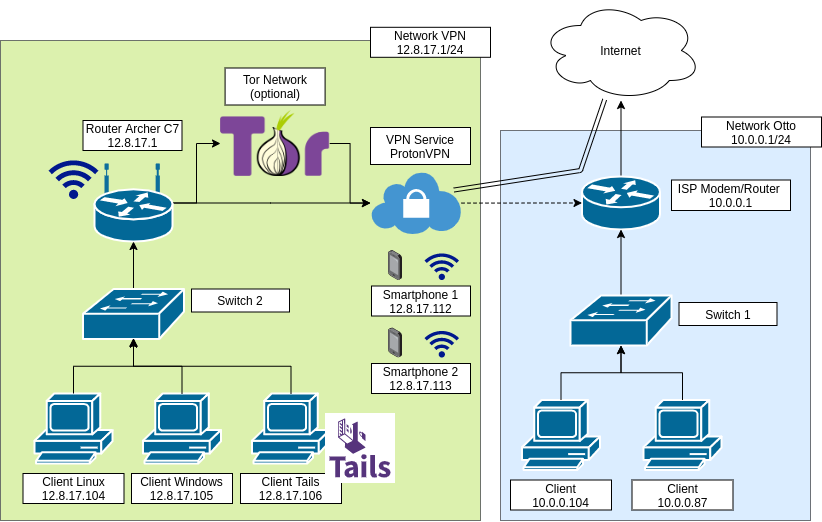}
    \caption{Displays the desired network setup including the chosen VPN service provider.}
    \label{pic:network_setup}
\end{figure}
\begin{figure}[h]
    \centering
    \includegraphics[width=0.5\columnwidth]{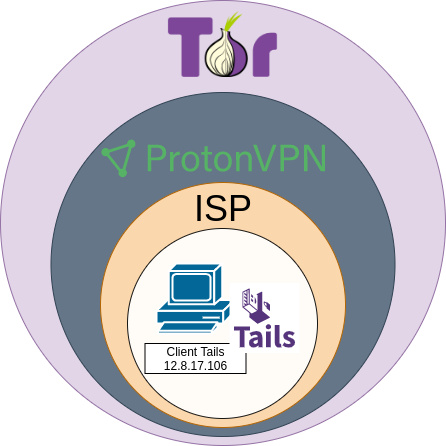}
    \caption{A simplification of the setup used and shown in Figure~\ref{pic:network_setup} in order to gain and preserve anonymity.}
    \label{pic:AnonymitySetup}
\end{figure}

In order to achieve such a disguise, we decided to create local layers, as seen in Figure~\ref{pic:AnonymitySetup}. We eliminated potentially identifying pieces of information lurking on the device used by removing the internal storage disk and booting the operating system ``Tails'' \cite{tails:2020} entirely from a USB-live stick. Before accessing any internet services, a virtual private network (VPN) was installed on the router, allowing us to hide our traffic from the ISP. In the final step, we accessed the TOR network through our VPN connection which allowed us to hide traffic from the VPN provider. Such setup is visually displayed in Figure~\ref{pic:network_setup}. In addition, Open-VPN software provides a functionality with the name ``remote-random'' where a new random server in an arbitrary country is selected during the boot-up of the router. In extreme cases, our approach can be further strengthened as free public internet access might be used where the VPN needs to be established on the mobile computer. However, in this case, surveillance cameras might be used to detect our identity, hence need to be strictly avoided. \cite{kutschera:2021}

\subsection{Interviews \& Statements}\label{sec:interviews}
The following expert interviews were made in alignment with the presence on social media or the profession.

\subsubsection{Interview, Scott Helme}\label{subsubsec:inthelme}
Mr.\ Scott~Helme, a security researcher, entrepreneur, and international speaker who provides training on hacking and encryption, was available to answer questions regarding findings from his Twitter and LinkedIn accounts. A post where the reflection on his car revealed details of his renovated home ultimately allowed findings such as his full home address, complete birthdate, detailed floor planning, and the price and date of the alleged purchase within a 40 min search. 
Mr.~Helme confirmed the findings and was resigned to legislation in the UK that required such information to be made public. He is nonetheless well aware that such information can be used to plan a burglary, to impersonate him or to produce personalized fraud letters sent via mail directly to his home. He also said the effort to hide or remove his data appeared to him as unbearable. \cite{kutschera:2021}

\subsubsection{Interview, Dr. Ries Bouwman}\label{subsubsec:riesbouwman}
International entrepreneur and co-founder of ``Omi's Apfelstrudel'' was asked for his view on privacy for SMEs. He was aware of the information that had been found and was happy that it was easily available to the public as none of the data represents a threat, but rather an opportunity for business partners to get in touch with him. Moreover, he uses public information to track if he or his business partners previously marked a new business as untrustworthy. \cite{kutschera:2021}

\subsubsection{Interview, Dr. Vesna Krnjic}\label{subsubsec:vesnakrnjic}
Dr. Vesna Krnjic, a researcher for privacy and security, at the Institute of Software Technology at the Graz University of Technology, was not surprised by the information found after an OSINT search of 1 hour and 50 minutes among her social media profiles. However, what she found most interesting was the approach used. Dr.~Krnjic fully supports the proposed change of law\footnote{See Section~\ref{sec:changeoflaw} \cite{kutschera:2021}}, as she sees no need for SMEs without direct contact to customers to publicly disclose the company address when it matches the private home address of the owner. In addition, Dr.~Krnjic believes the mobile phone signature offers the infrastructure needed for easily supporting an additional layer of security for such vulnerable information, as the current situation allows data gathering without registration or limitation. Ultimately, Dr. Krnjic points out that in today's internet life, anyone and everyone might potentially reveal personal information about themselves without being aware of it. Regarding the lack of consciousness about these issues in the broad mass of internet users, she feels this is a little frightening when she reflects on the information that can be revealed by photos posted online, especially when considering the involvement of artificial intelligence or the use of unlimited time for analysis. \cite{kutschera:2021}

\subsubsection{Interview, Austrian SME entrepreneur}\label{subsubsec:invsme}
The Austrian SME Entrepreneur\footnote{\hspace{0.1cm}Due to privacy concerns, it was
decided to redact personal information.}, who owns several homepages and hosts her podcast, states that social media plays a crucial part in her daily business. Travel enthusiasts but also professional scuba divers appreciate her expertise the most. The found data was already known to be publicly available; however, the approach used was new to her. For her, the obligation within Austria's legislation to publish an imprint is most polarizing as it is helpful but also concerning at the same time. Her measures against criminals or attacks include but are not limited to: Never posting pictures of a destination that she is currently reviewing; Posts are published according to a different time zone to prevent information leakage on observed posting times; Legally binding her business partners not to release any statements or information revealing her current or future locations. \cite{kutschera:2021}

\subsubsection{Interview, Henry from Techlore}\label{subsubsec:techlore}
Henry, the owner of Techlore, an organization in the United States that provides content and consultation on privacy and security, states that physical danger is the most likely threat behind publicly available data, which can be compensated with local security protections. He has mixed feelings regarding the obligation of publicly available company information or another protection layer. In Henry's vision, education and the availability of alternative possibilities is a solution to the problem. Hence make the general public and business owners aware of security and privacy threats and, in addition, giving easy possibilities to implement proper measures such as creating businesses outside one's home address. \cite{kutschera:2021}

\subsubsection{Interview, Heimo Flechl}\label{subsubsec:intflechl}
Mr.~Flechl states that OSINT can not be declared as criminal activity per se but certainly will be used for further criminal activities. The monetary value varies from subject to subject and may have an additional impact on integrity or reputation. However, damages can also occur on a physical or psychological level.
With this in mind, Mr. Flechl reminds us that arbitrarily collected data may not be of value on the black market but, may well be crime as a service. For example, crime as a service could be in the form that one criminal creates and sells a Trojan that collects banking data from victims and maybe infects contacts stored on the device of the victim. The collected data is then bought and used by another criminal who exploits the information and with the possible involvement of other criminals who are able to perform money laundering. \cite{kutschera:2021}

For Mr.~Flechl, the imprint of companies is one piece of information, but it is certainly not enough to gain the trust of employees in order to perform a successful fraud. A question that must be asked is whether the information that is given on the company page is necessary for fulfilling the legal or business requirements and to gain the trust of customers. This is also the case for individual persons who are sharing information on various social media platforms, when a closer look at the primary subject of the posting might reveal minor yet still visible and revealing details such as street names, house numbers, famous buildings, prominent landmarks and security measures, but also even the absence of such details. \cite{kutschera:2021}

\subsubsection{Conclusion of the interviews Sec.\ref{subsubsec:inthelme} - \ref{subsubsec:intflechl} }
Summarizing of the interview contents has emphasized that privacy is a polarizing topic. On the one hand, entrepreneurs use publicly available business information to verify potential partners' integrity and decide whether to conduct business with them, as the interviews with Dr. Bouwmann and the Austrian SME Entrepreneur have shown. On the other hand, the expertise of Mr. Flechl shows how criminals can misuse this information.

\section{Evaluation} \label{ch:evaluation}

\subsection{K-index table}\label{subsec:k-index-table}
We found no method or formula that satisfied our requirement for making incidental data comparable and to correlate it with one's broadcast range. This is the ultimate reason behind our approach of the K-index. Stated simply, the \emph{K-index} is a weighted sum of different categories on incidental data where in one case the value is capped. At the same time, the \emph{Relative K-index} is set in relation to the broadcast range of the person. However, we had issues implementing caped values as the PostgreSQL database used limited us in using only mathematical operations and no if statements within a query, which led to the adapted limitation equation shown in Equation~\ref{eq:equation_kindex}. 

\begin{figure}[H]
    \centering
    \includegraphics[width=\columnwidth]{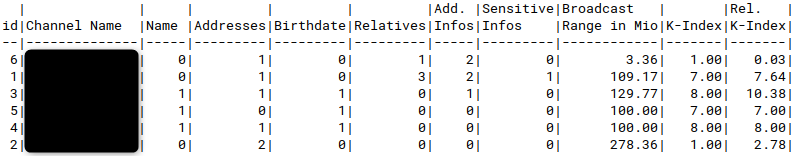}
    \caption{K-Index table exported and redacted from the dedicated relational database PostgreSQL.}
    \label{fig:k_index}
\end{figure}

\begin{table}[H]
    \centering
    \begin{tabular}{l|l}
    \hline
    \rowcolor{gray5}k-array          & weight \\
    \hline
    k.name           & 3   \\
    k.birthdate      & 4   \\
    k.relatives      & 0,5 \\
    k.add.info       & 0   \\
    k.sensitive.info & 5   
    \end{tabular}
\end{table}

\begin{equation} \label{eq:equation_kindex}
k.index = \frac{2\cdot 1}{1+e^{-20\cdot (k.addresses)}}-1 \pmod{2} + \sum_{i=1}^{n} k_i x_i 
\end{equation}

\begin{equation} \label{eq:equation_relkindex}
rel.k.index = \frac{broadcast.range \cdot 0,000001}{100} \cdot k.index
\end{equation}

\subsection{Mock-up case Austria-Austria}\label{sec:MUC_Austria} 
The gathering and extraction of incidental data from social media platforms clearly raises legal questions. With this in mind, we attempted to summarize the most relevant facts and created a fictional case involving two Austrian citizens, where the plaintiff is the owner of social media accounts from which incidental data has been extracted, and to show what legal aspects need to be taken into consideration in such a case.

\subsubsection{Facts}\label{subsec:muc_facts_austria} 
In Table~\ref{tab:plaintiff_social_media_accounts} the plaintiff's social media accounts are described.

\begin{table}[H]
    \caption{List and properties of the owned accounts by the plaintiff.}
    \begin{tabular}{l|lll}
        Platform & Public Account Name & Media Reach & Content\\
        \hline
        YouTube                  & MU Account YT & 950.000  & 450 videos\\
        Twitter                  & MU Account TW & 60.000  & 1.800 tweets\\
        Facebook                 & MU Account FB & 60.000  & unknown\\
        Reddit                  & MU Account RD & unknown & 1.000 comments\\
        \hline
    \end{tabular}
    \label{tab:plaintiff_social_media_accounts}
\end{table}

The plaintiff owns several social media accounts and has developed his business from a simple hobby to a legal entity that was established in 2014. Due to financial limitations and non-existing physical customer contacts, the plaintiff decided to use a separate room in his own home for his business. The plaintiff uploaded a 15:20 minute video with the title ``My beautiful tomatoes are growing so well'' onto the video platform YouTube in January 2019. The video had 12,082,017 views. The garden in the back of the plaintiff's home is intentionally visualized in the video. However, unintentionally the video shows two sides of the plaintiff's house, including its colour and structure, as well as a five-floor tall building and mountains with specific patterns in the vicinity. The plaintiff also mentions issues with stalkers in the past, which following a newspaper interview, led to the given name only being published, instead of the full name.\par

The defendant accessed the aforementioned video and created screenshots at 1:25, 5:55, and 13:12 min.
The second attempt to geolocate the address of the plaintiff was successful after the name of the channel was typed into the Google search engine and ultimately hinted at by the auto-completion. 
The hinted full name a query against the Austrian Business Licence Information System (GISA\footnote{From German language ``Gewerbeinformationssystem Austria – GISA''}) resulted in a hit that was further verified using VPL. The address was then for a cross-check entered into the public land register of Styria (GIS Kataster\footnote{From German language ``Geo Informations System Kataster Steiermark (GIS Kataster)''}), where the property owner's name matched the alleged full name of the plaintiff. 
As a result, the defendant assumed that the information found on the address, full name, date of birth, and owned property of the plaintiff was accurate and cross-confirmed throughout several publicly available databases.

\subsubsection{Legal obligations}\label{subsec:muc_legalobligations_austria} 
\paragraph{YouTube}
In accordance with terms of service\footnote{Note that the legally binding version of 22\textsuperscript{nd} of July 2019 is written in German.} (TOS) the uploading user is still responsible for the content. However, the viewing users have restrictions as stated in the TOC as follows \emph{``The user is not allowed to access, reproduce, download, distribute, transmit, display, sell, license, modify, adapt, or otherwise use any portion of the service or content except in the manner permitted by the service.''} \cite{ytterms:2019}

\paragraph{Google Search Engine}
The TOS of the Google search engine is not per se violated by hinting at a full name. However, if the search results are used in a violation of applicable law, TOS would inevitably be violated as well. However, if the defendant were to use the search results to violate applicable law, for instance by stalking or harassing the plaintiff, a violation of the Terms of Service would be inevitable. \cite{googleterms:2020}

\paragraph{Gewerbeinformationssystem Austria (GISA)}\label{subsubsec:gewo1994}
Austrian law §~365e Gewerbeordnung 1994 (Trade Code) ('GewO 1994'), BGBl I 45/2018, grants the unrestricted  access to GISA. However, information as specified in §~365a Sec.2 Fig.1-8 GewO 1994, among others, is the private home address and shall only be provided if the requesting person substantiates a legitimate interest in the information. 

\paragraph{Public land register (``GIS Kataster'')}
A sudden deactivation of the public land register service ``GIS Kataster'' was not covered by a recent change of legislation. However, a request for comment was issued to the Styrian Provincial Government and answered as follows: \emph{``[...] Due to the data protection regulation and increased inquiries, the constitutional service of the province re-evaluated the question of access to land register data via the GIS Styria. The land register access continues to be available to the public on a limited basis via the other known platforms. [...]''}. \cite{kutschera:2021}

\paragraph{Regulation EU 2016/679 GDPR \& DSG}
As the defendant, as well as the plaintiff, are citizens of Austria, the Regulation EU 2016/679 General Data Protection Regulation (GDPR) is inter alia Art.~3 applicable. Terms, such as \textit{personal data} and \textit{processing}, used within Regulation EU 2016/679 GDPR are defined within Art.~4.
\emph{Personal data} relates to any identifiable or identified information of a natural person.
Whereas operations either manually or automatically within a set of operations or a single operation performed on personal data is defined as \emph{processing}. Such processing operations are defined by Art.~4 Sec.~2 of Regulation EU 2016/679 GDPR, as the collection, recording, organization, structuring, storage, adaptation or alteration, retrieval, consultation, use, disclosure by transmission, dissemination or otherwise making available, alignment or combination, restriction, erasure or destruction of personal data. \cite{unger:2018}
Art.~15 Sec.~1 Regulation EU 2016/679 GDPR grants a data subject the right to ask for confirmation and access to personal data in possession of the data controller from the data controller. Inter alia, Art.~2 Sec. 2 Lit.\ c Regulation EU 2016/679 GDPR, grants an exception in case of \textit{``a natural person in the course of a purely personal or household activity''}. \cite{unger:2018} Moreover, based on Regulation EU 2016/679 GDPR, the ruling of Oberster Gerichtshof (Austria) OGH 6 Ob 150/19f
refers in section 5 of the reasoning inter alia the decision of European Court of Justice (ECJ) C-212/13, where the latter is based on the predecessor of Regulation EU 2016/679 GDPR, namely Directive 95/46/EC. As a matter of fact, the ruling of Oberster Gerichtshof (Austria) OGH 6 Ob 150/19f, refers in section 5 of the reasoning inter alia to the decision ECJ C-212/13, where the principle of household activity cannot be claimed if the gathered personal data is also used in order to protect property. This jurisdiction makes clear that household activity has a restricted meaning.
In view of the aforementioned, two facts can be drawn. Firstly, ECJ rulings based on Directive 95/46/EC can still be used to help interpret the successor Regulation EU 2016/679 GDPR. Secondly, a natural person that does not solely process data based on private and household activity, which needs to be viewed as restricted, also needs to  apply with Regulation EU 2016/679 GDPR.\par 
This ultimately implies that the defendant (Section~\ref{subsec:muc_facts_austria}) needs to be viewed as a controller and eventually shall grant rights and protect personal data to the data subject as per Regulation EU 2016/679 GDPR.

\section{Results} \label{ch:results}

\subsection{Incidental data}\label{sec:awareness}
As shown in Section~\ref{subsec:gatehringprocessyoutube}, \ref{sec:troyhuntmanualgather} and \ref{subsubsec:vesnakrnjic} we were able to find incidental data among a small number of social media accounts. Given this fact, we need to reject our \emph{Null-Hypothesis} from Section~\ref{sec:hypothesis} and confirm the hypothesis that social media postings may include incidental data which, in addition, may have the potential to cause harm.  
Incidental data was found among entrepreneurs, security experts, politicians, and other individuals with various specializations. \cite{kutschera:2021,standardkurz:2019} Among the incidental data, information such as private addresses, detailed floor-plan, price of the property respectively buildings or flats, security measures, and information on alleged relatives was found. In the case where we found incidental data about the former Chancellor of Austria, who was in office at the time of the analysis, the Austrian Federal Chancellery was constructively involved at an early stage of our findings. 
Further, we found victims of stalking that remained on social media platforms and showed more awareness towards privacy. Sadly, it was still possible to gather incidental data from content posted after the incidents, such as the private address, detailed information on the main home- and other properties, as well as information on relatives. 
Equally important is the role of public- and social-watchdogs. However, case law has shown that individuals well known to the public must allow private information to be (re)published under certain circumstances. In this light, gathered incidental data from an online security expert may not withstand the balancing between Art.8 ECHR and Art.10 ECHR.

\subsection{Methods for incidental data }\label{subsec:summarized-security-measures}
\begin{figure}[H]
    \centering
    \includegraphics[width=\columnwidth]{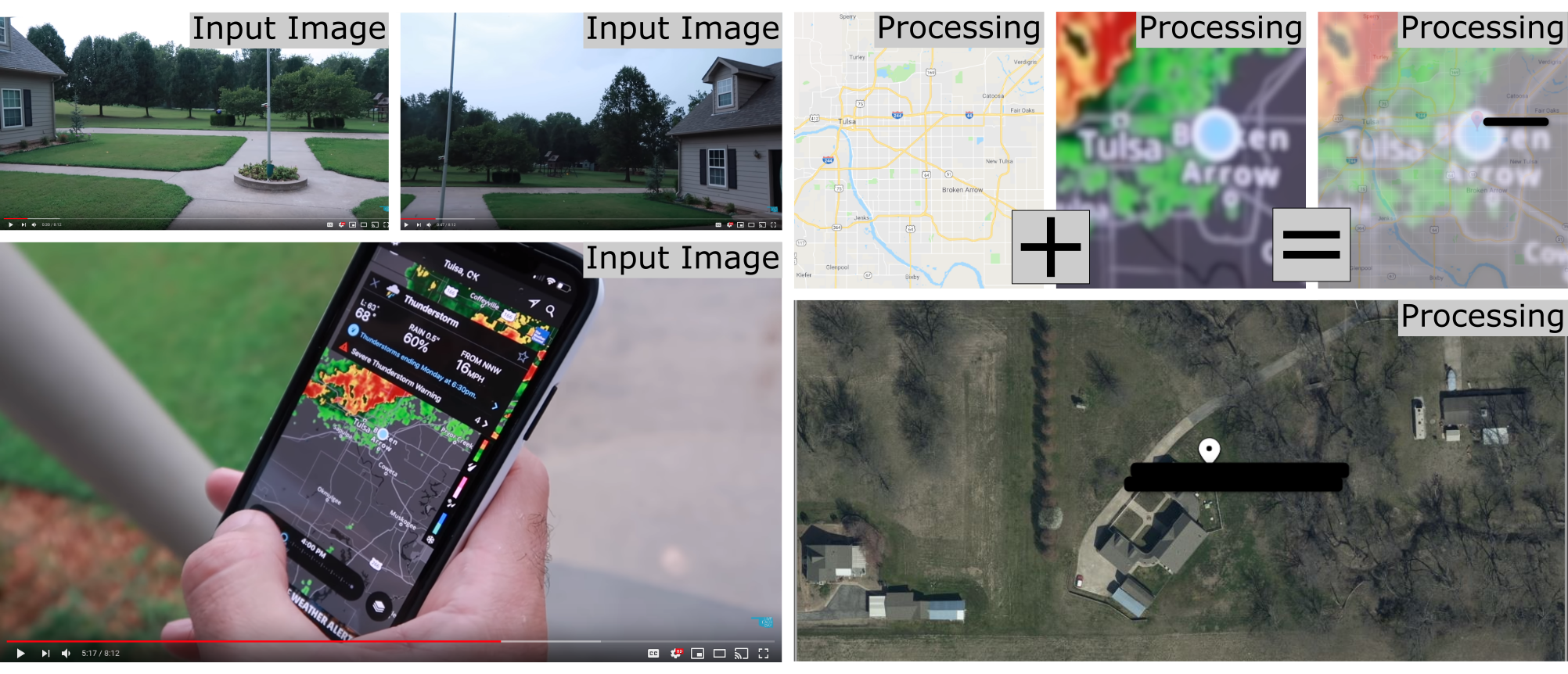}
    \caption{Images from a social media post that were used and processed to identify a location. \cite{googlemapsoverall:2021,youtubeoverall:2021}}
    \label{fig:guide_locationidentification}
\end{figure}

\begin{figure}[H]
    \centering
    \includegraphics[width=\columnwidth]{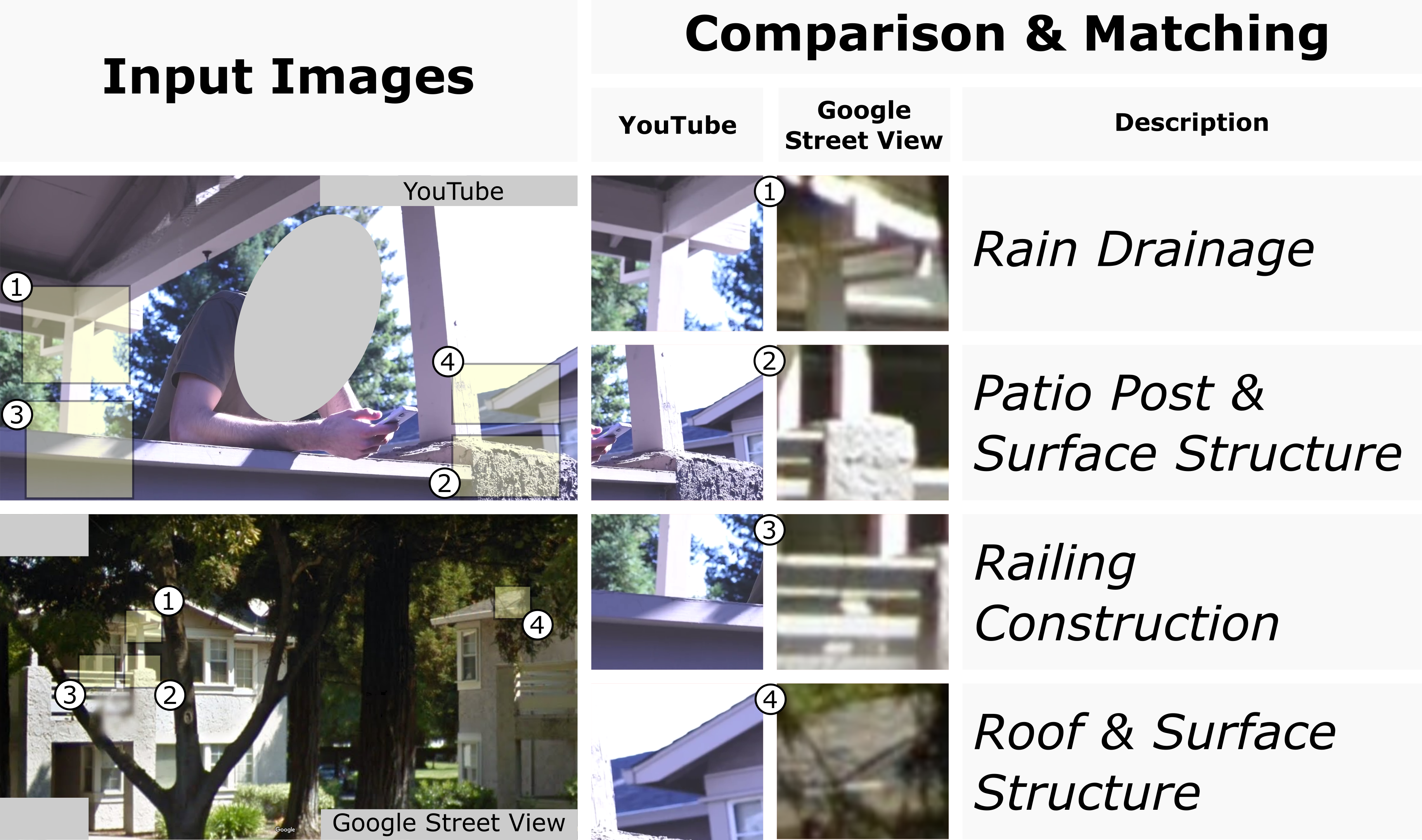}
    \caption{Matching features from a social media post compared with Google Street View to verify an address. \cite{googlemapsoverall:2021,youtubeoverall:2021}}
    \label{fig:guide_locationverification}
\end{figure}

We found two methods most useful. First, the \emph{identification of a physical location} (IPL) where the address of a person is not known but due to an OSINT investigation found with a high level of certainty. The IPL-method is visualized in Figure~\ref{fig:guide_locationidentification}. Secondly, \emph{verification of a physical location} (VPL) where the address might be known but not confirmed or the information is outdated. VPL-method is used to verify the address of a person with high certainty, as visualized in Figure~\ref{fig:guide_locationverification}. 

\subsection{Summarized awareness guideline}\label{sec:secmeasures}
Given our observation and expert interviews, we concluded the following best practice measures to improve one's online-security and -privacy behavior. We acknowledge that some measures are not suitable for every person and want to emphasize that the measures might also strengthen the privacy of others if later published content was taken from the property of someone else.   

\begin{enumerate}
    \item Avoid posting content that includes house numbers or street names.
    \item Be on the lookout for reflections in mirrors as well as on surfaces such as cars, windows, vitrines, glasses, sunglasses, or watches.
    \item Post content of vacations - if at all - only after the vacation has ended.
    \item Avoid repetition of vacations or periods of absence, such as ``during New Years Eve I am - always - on a one-week trip''.
    \item Posts should be in accordance with a single time zone irrespective of a current and temporary time zone.
    \item Avoid posting any information from parcels or letters, such as tracking number, full address, names, or QR codes.
    \item Do not post IDs such as driver's license, personal ID, credit- or debit card, et cetera.
    \item Avoid posting scenes that include location-based map materials, such as navigation maps, weather- or fitness apps, et cetera.
    \item Close all curtains or post content where no windows are visible.
    \item Try tilting the camera angle as low as possible when showing the own property.
    \item Be aware that shadows or the sun's position can also hint at additional information about the location.
    \item Do not share fitness routes that start or end at your home location.
    \item Don't share information of your own or surrounding WLAN/WiFi SSID's
\end{enumerate}

\subsection{Proposed change of law}\label{sec:changeoflaw}
Implementing a change of law as stated below \footnote{Since the Austrian legislation is in force in its entirety only in the German language, we decided to present the proposed change of law also in German. This decision was driven by the desire to make an implementation of the proposed change as effortless as possible.}  would increase the right to respect for private and family life (Art.8 ECHR) of entrepreneurs and their families without infringing against the rights of other parties. This protection can be archived by accessing the currently unrestrictedly accessible information by restrictions imposed through the qualified electronic signature as with Regulation EU 910/2014 (e-IDAS). Moreover, this would ensure that a private address will have a higher standard of protection as indicated by  § 365a (2) 4 GewO 1994.\\

\noindent
\textit{``Die Gewerbeordnung 1994 – GewO 1994, BGBl. Nr. 194/1994, zuletzt geändert durch das
Bundesgesetz BGBl. I Nr. 65/2020, möge wie folgt geändert werden:\\
Dem §365e wird folgender Abs.6 angefügt:\\
(6) Ist die Wohnanschrift (§365a Abs. 2 Z 4) dieselbe wie der Standort der Gewerbeberechtigung (§365b Abs. 1 Z 3), hat die Behörde den Standort der Gewerbeberechtigung gleich wie seine Wohnanschrift (§365a Abs. 2 Z4) zu behandeln und nur darüber Auskunft zu erteilen, wenn der Auskunftswerber ein berechtigtes Interesse an der Auskunft glaubhaft macht.
\begin{enumerate}
    \item Der Gewerbetreibende hat jederzeit die Möglichkeit der Behörde formlos mitzuteilen, den Standort seiner Gewerbeberechtigung (§365b Abs. 1 Z 3) jedermann, auch ohne Glaubhaftmachung von berechtigtem Interesse, zu beauskunften.
    \item Die Mitteilung über §365e Abs. 6 Z 1 an die Behörde kann mündlich, telefonisch, telegrafisch, schriftlich, fernschriftlich, mit Telefax, im Wege automationsunterstützter Datenübertragung oder in jeder anderen technisch möglichen Weise erfolgen.''
\end{enumerate}
}

Into English translated version of the proposed change of law:\\
\noindent
\textit{``The Gewerbeordnung 1994 - GewO 1994, Federal Law Gazette No. 194/1994, as last amended by the Federal Law Gazette I 65/2020, shall be amended as follows:\\
The following Paragraph 6 shall be added to §365e:\\
(6) If the residential address (Section 365a (2) 4) is the same as the location of the business license (Section 365b(1)(3)), the authority shall treat the location of the business license in the same way as its residential address (Section 365a (2) 4) and shall only provide information about it if the person requesting the information substantiates a legitimate interest in the information.
\begin{enumerate}
    \item The trader has at any time the possibility of informing the authority informally, the location of its trade license (§365b (1) 3) to provide information to anyone, even without credible justified interest.
    \item The notification of §365e (6) 1 to the authority may be made orally, by telephone, telegraph, in writing, by telex, by fax, by means of computer-assisted data transmission or in any other technically feasible manner.''
\end{enumerate}
}
\section{Conclusion and Outlook}\label{ch:conclusion}

\subsection{Conclusion}\label{sec:conclusion}
All things considered, incidental data can be found among people with different personal and professional backgrounds. Interestingly, expert interviews have shown that publicly available information is a polarizing topic among entrepreneurs as it can be an ominent threat to one's private life, or by contrast it may enhance and secure one's own business. Controversial is the finding of the pre-announced and updated trip of several thousand kilometres as an online security expert should lead as a role model.

\subsection{Outlook}
Despite our tendency to consider incidental data as an awareness problem that can be solved by education, the implementation of a tool that is able to detect critical content in a posting and warns a user before the post is made public, would be of substantial value. At the same time, tools of this kind also have the potential to be used against people and consequently raise ethical issues. The problems currently surrounding incidental data might eventually be solved by interdisciplinary measures such as education and awareness programs and furthermore be strengthened by legislation as we proposed in our presented change of law \cite{kutschera:2021}.


\begin{acknowledgements}
Special thanks are due to Michael~Brickmann, Dr.~Sabine~Proßnegg further also Dr.~Bernd~Pichlmayer for their helpful input and for many hours of fruitful discussions.
\end{acknowledgements}

%
%

\bibliographystyle{spmpsci}      
\bibliography{references}   


\end{document}